\documentstyle[ApJ,times,PSfig]{article}

\def\bdm{\begin{displaymath}} \def\edm{\end{displaymath}}
\def\beq{\begin{equation}} \def\eeq{\end{equation}}
\def\ie{i.e$.$~} \def\eg{e.g$.$~} \def\etal{et al$.$~}
\def\eq{eq$.$~} \def\eqs{eqs$.$~} 
\def\epsel{\varepsilon_e}  \def\epsmag{\varepsilon_B}
\def\cm3{\;{\rm cm^{-3}}} \def\deg{^{\rm o}} \def\muas{\,{\rm \mu as}}

\def\simg{\mathrel{%
      \rlap{\raise 0.511ex \hbox{$>$}}{\lower 0.511ex \hbox{$\sim$}}}}
\def\siml{\mathrel{%
      \rlap{\raise 0.511ex \hbox{$<$}}{\lower 0.511ex \hbox{$\sim$}}}}

\begin{document}

\title{A Jet Model With A Hard Electron Distribution for the Afterglow of GRB 000301c}

\author{A. Panaitescu}
\affil{Dept. of Astrophysical Sciences, Princeton University, Princeton, NJ 08544}

\begin{abstract}

 The parameters of the GRB 000301c afterglow are determined within the usual framework 
of synchrotron emission from relativistic ejecta, through fits to the available 
radio-to-optical data. It is found that $1)$ the jet energy after the GRB phase is
$\sim 3\times 10^{50}$ erg, $2)$ the initial opening angle of the jet is $\sim 13 \deg$,
$3)$ the medium that decelerates the afterglow has a density $\sim 25\;{\rm cm^{-3}}$, 
$4)$ the power-law distribution of the shock-energized electrons is hard, with an index
around 1.5, and $5)$ the cooling frequency was located below the optical domain. 
Furthermore we find that the collimation of ejecta alone cannot explain the sharp and 
large magnitude break observed in the $R$-band emission of this afterglow at few days, 
and that a high energy break in the electron distribution, corresponding to an
electron energy close to equipartition, together with the lateral spreading of the jet 
accommodate better this break. Microlensing by a star in an intervening galaxy produces
only a flattening of the afterglow emission and cannot explain the mild brightening 
exhibited by the optical emission of 000301c at $\sim 4$ days. 
 
\end{abstract}

\keywords{gamma-rays: bursts - ISM: jets and outflows - methods: numerical -
          radiation mechanisms: non-thermal - shock waves}

\section{Introduction}

 The afterglow of the GRB 000301c is one of the best observed to date, its emission having
been monitored over 1.5 decades in time at radio frequencies (Berger \etal 2000) and in the
optical (Bhargavi \& Cowsik 2000, Masetti \etal 2000, Sagar \etal 2000, Jensen \etal 2001,
Rhoads \& Fruchter 2001). Furthermore, 000301c is one of three afterglows whose optical 
emission decay exhibited a strong steepening (break) at few days. The steepening follows
a flattening, or mild brightening, in the optical light-curve. The break in the optical 
emission was attributed to a jet geometry by Berger \etal (2000), while the relatively 
short-lived brightening at 3--5 days was interpreted by Garnavich, Loeb \& Stanek (2000) 
as evidence for gravitational microlensing due to a star in a intervening galaxy. Rhoads
\& Fruchter (2001) point out that the sharpness of the optical break poses difficulties for
a jet interpretation.

 In this work we model the emission of 000301c in the framework of relativistic ejecta 
energizing the swept-up gas surrounding the GRB source (M\'esz\'aros \& Rees 1997). 
Our aims are to assess the ability of the jet model to accommodate the behaviors exhibited 
by the radio-to-optical emission of 000301c and to determine its basic physical parameters 
(jet energy, initial opening, surrounding medium density, etc). Furthermore, by including 
microlensing in our model, we also test its ability to reproduce the optical brightening 
seen around 3.5 days.

\section{Model Description}

\subsection{Afterglow Dynamics and Synchrotron Emission}
\label{synchro}

 The jet dynamics is given by a set of differential equations (Panaitescu \& Kumar 2001) for the 
Lorentz factor $\Gamma$, mass, and aperture $\theta$ of the jet as functions of radius $r$. 
The jet lateral spreading is negligible in the early phase of the expansion. In the adiabatic 
limit, the jet Lorentz factor evolves as $\Gamma \propto r^{-3/2} \propto t^{-3/8}$ for a uniform 
external medium, where $t$ is the observer time, and as $\Gamma \propto r^{-1/2} \propto t^{-1/4}$ 
for a medium whose density decreases as $n(r) \propto r^{-2}$ (\eg M\'esz\'aros, Rees \& Wijers 
1999), \ie for a wind ejected at constant speed before the release of the relativistic jet. 
The lateral spreading increases the jet sweeping area and alters significantly its dynamics when 
$\Gamma$ drops below $\theta^{-1}$ (Rhoads 1999), the jet deceleration becoming exponential, 
$\Gamma \propto e^{-kr} \propto t^{-1/2}$, as long as the jet is sufficiently relativistic 
($\Gamma \simg 10$). 

 We calculate numerically the jet dynamics, taking into account the energy loss through emission 
of radiation. The jet is assumed to have sharp boundaries and to be uniform, with no angular 
gradients (of energy or mass) within its aperture. From the dynamics of the jet, one can calculate 
the properties of the shocked fluid (density, magnetic field, typical electron random Lorentz factor), 
which are necessary for the computation of the synchrotron emission.

 We assume that the shock injects in the downstream region relativistic electrons with a 
power-law distribution, and approximate the distribution of electrons resulting from injection 
and cooling as a dual power-law (Sari, Piran \& Narayan 1998), with a minimum and a break set by
the injected ($\gamma_i$) and cooling ($\gamma_c$) electron Lorentz factors. 
This approximation allows some analytical results for the self-absorption frequency, Compton 
parameter, radiative losses, and peak synchrotron flux to be implemented in the code, avoiding
thus time-consuming integrations over the electron distribution. We have tested the accuracy 
of this approximation and find that it is very good if radiative losses are small. For significant
radiative losses (above 50\%) it may mis-estimate the electron distribution and $\gamma_c$ by 
a factor $\siml 3$, thus the model fluxes in some frequency ranges at for certain durations
may be off by a similar factor.

 In the calculation of the afterglow emission we allow for the existence of a high energy break 
at $\gamma_*$ in the electron distribution, at which the shock-acceleration becomes inefficient. 
This break could be due to the escape of particles from the acceleration region and to radiative 
cooling during acceleration, and becomes important if the emission from $\gamma_*$-electrons 
crosses an observational band. In this case the resulting afterglow emission fall-off depends 
on the exact shape of the electron distribution at and above $\gamma_*$. For simplicity we shall 
assume that at $\gamma_*$ there is a sharp transition to a steeper power-law.

 The afterglow spectrum has three breaks: the synchrotron characteristic frequencies $\nu_i$ 
and $\nu_c$ at which the $\gamma_i$- and $\gamma_c$-electrons radiate, and the self-absorption 
frequency $\nu_a$. Between these breaks the spectrum is a power-law of an index dependent on 
the ordering of the breaks (\eg Sari \etal 1998). The steepening of the electron distribution 
at $\gamma_*$ adds another break, at the frequency $\nu_*$ at which the $\gamma_*$-electrons 
radiate. As the jet is decelerated, these four frequencies evolve, the passage of one of them 
through an observing band yielding a chromatic steepening of the afterglow emission fall-off. 
Furthermore an achromatic steepening of the afterglow decay is expected around the time when,
due to the decrease of the relativistic beaming of the afterglow emission, the jet edge becomes
visible to the observer (Rhoads 1999).

 In our treatment, the afterglow model has 8 parameters. Three of them give the jet dynamics: \\
\hspace*{2mm} $1)\;\; E_0$, the jet energy at the end of the GRB phase\\
\hspace*{2mm} $2)\;\; \theta_0$, the initial jet aperture (half-angle) \\
\hspace*{2mm} $3)\; n$, the particle density of the uniform external medium or the parameter  
  $A = (1/4\pi) (\dot{M}/m_p v)$ for a medium with a density profile 
  $n(r) = A r^{-2}$, resulting from the ejection by the GRB progenitor of a wind with
  mass loss rate $\stackrel{.}{M}$ and constant speed $v$. 
The other five are parameters for the microphysics of shock acceleration and generation of
magnetic fields, and determine the spectrum of the emergent synchrotron emission: \\
\hspace*{2mm} $4)\;\; \epsmag < 1/2$, the ratio of the magnetic energy density to the energy 
   density of the freshly shocked fluid. \\
\hspace*{2mm} $5)\;\; \epsel \ll 1$, the ratio of the energy density of newly accelerated electrons, 
   if they all had the minimum Lorentz factor $\gamma_i$, to the energy density of the post-shock 
   gas. This parameter quantifies $\gamma_i$. \\
\hspace*{2mm} $6)\;\; p$, the index of the power-law distribution ${\cal N}_i (\gamma) \propto 
   \gamma^{-p}$ (for $\gamma_i < \gamma$) of the injected electrons. \\
\hspace*{2mm} $7)\;\; \epsilon < 1/2$ (greater than $\epsel$), the fractional internal energy stored 
   in electrons. This parameter determines the high-energy break $\gamma_*$ of the electron 
   distribution for $p < 2$ and any $\epsel < \epsilon$ or for $p > 2$ and $\epsilon < [(p-1)/(p-2)] 
   \epsel$. \\
\hspace*{2mm} $8)\;\; \delta p$, the steepening of electron distribution index above $\gamma_*$. 
A quantitative description of these five parameters can be found in Panaitescu \& Kumar (2001).

 We note that, if the jet evolution is quasi-adiabatic, then its dynamics and the afterglow
emission are independent on $\Gamma_0$. If radiative losses are significant, the afterglow 
emission depends on $\Gamma_0$ because the magnitude and evolution of the radiative losses is
$\Gamma_0$-dependent. However, this dependence is rather weak, so that the data cannot be used
to obtain meaningful constraints on the initial jet Lorentz factor. For this reason we shall keep 
it fixed: $\Gamma_0 = 500$.

 In the calculation of the radiative losses and cooling frequency we take into account synchrotron 
self-absorption and inverse Compton scatterings (which may be dominant if the magnetic field is 
sufficiently weak). The received synchrotron flux is calculated by integrating it over the jet 
evolution, taking into account the effect of the source relativistic motion and spherical curvature 
on the photon arrival time and energy, but ignoring the light shell-crossing time. The observer is 
assumed to lie on the jet axis.

\subsection{Interstellar Scintillation}
\label{iss}

 Fluctuations on a short timescale of the radio emission of GRB afterglows, caused by 
inhomogeneities in the Galactic interstellar medium have been predicted by Goodman (1997). 
As shown by Berger \etal (2000), the well-sampled 8.46 GHz emission of 000301c exhibits 
such fluctuations. We include the effect of the interstellar scintillation (ISS) by adding 
in quadrature to the reported errors the expected amplitude of the model flux fluctuations due 
to ISS. The calculation of the modulation amplitude follows the treatment given by Walker (1998),
and is summarized below.

 At a frequency $\nu$  below a certain critical value $\nu_0$, scintillation occurs in the 
strong regime. {\sl Diffractive} scintillation (DS) is caused by the interference of light 
coming from many patches of coherence within the "refractive" disk of size $\theta_r \equiv 
\theta_{F0} (\nu_0/\nu)^{11/5}$, where $\theta_{F0}$ is the angular size of the first Fresnel 
zone at the critical frequency $\nu_0$. DS has amplitude $m_d^{(ps)} = 1$ if the source is 
point-like (\ie its size $\theta_s$ is smaller than $\theta_d \equiv \theta_{F0} [\nu/\nu_0]^{6/5}$) 
and if the observing window is narrower than the decorrelation bandwidth $\Delta \nu = 
\nu (\nu/\nu_0)^{17/5}$. For an ``extended" source ($\theta_d < \theta_s$) the modulation index 
is reduced to $m_d^{(es)} = \theta_d/\theta_s$.

 {\sl Refractive} scintillation (RS) is due to inhomogeneities in the scattering screen on the 
refractive disk scale. In the strong scattering regime the RS amplitude is $m_r^{(ps)} = 
(\nu/\nu_0)^{17/30}$ if $\theta_s < \theta_r$, while for $\theta_r < \theta_s$ the modulation
index is $m_r^{(es)} = m_r^{(ps)} (\theta_r/\theta_s)^{7/6}$. At frequencies above $\nu_0$, 
RS occurs in the weak scattering regime and have amplitude $m_w^{(ps)} = (\nu_0/\nu)^{17/12}$ if 
$\theta_s < \theta_F$, where $\theta_F \equiv \theta_{F0} (\nu_0/\nu)^{1/2}$ is the size 
of the Fresnel zone. For $\theta_F < \theta_s$ the RS amplitude is reduced to $m_w^{(es)} =
m_w^{(ps)} (\theta_F/\theta_s)^{7/6}$.

 The amplitude of the radio ISS depends therefore on two properties of the interstellar 
medium, $\nu_0$ and $\theta_{F0}$, which depend on the scattering measure and distance to the 
scattering screen. The maps of Walker (1998) give $\theta_{F0} \sim 5.0 \muas$ and 
$\nu_0 \sim 4.4$ GHz in the direction of GRB 000301c, thus the measurements at the two lowest 
frequencies (4.86 GHz and 8.45 GHz) are close to the critical frequency. The above equations 
for the ISS amplitude are asymptotic results, strictly valid at frequencies far from $\nu_0$, 
and slightly overestimate the modulation amplitude at $\nu \sim \nu_0$. For this reason we set 
an upper limit of 50\% to the modulation index calculated with the asymptotic results above.

 The source size $\theta_s$, on which the ISS modulation may depend, can be easily calculated 
analytically at times before lateral expansion becomes important. During this phase the jet aperture 
$\theta$ exceeds $\Gamma^{-1}$, thus the observer receives photons mostly from a spherical cap of 
opening $\tilde{\theta} = \Gamma^{-1}$ around the jet axis. Therefore the size of the source seen 
by the observer is $\theta_s = r/\Gamma$. For a uniform medium ($\Gamma \propto r^{-3/2}$) it can 
be shown that a photon emitted by the fluid moving off-axis at an angle $\tilde{\theta}$ arrives 
at observer 5 times later than of a photon traveling along the jet axis. For a wind-like medium 
($\Gamma \propto r^{-1/2}$) the above factor is 3. Applying these correction factors to the 
expressions for the fireball radius given by Panaitescu \& Kumar (2000), which contain the arrival 
time of the photons emitted along the jet axis, we obtain
\beq
 \theta_s = 0.89\; D_{A,28}^{-1} (1+z)^{-5/8} ({\cal E}_{0,53}/n_0)^{1/8}\, t_d^{5/8} \;\muas 
\label{thetas0}
\eeq
for a homogeneous medium and 
\beq
 \theta_s = 0.60\; D_{A,28}^{-1} (1+z)^{-3/4} ({\cal E}_{0,53}/A_*)^{1/4}\, t_d^{3/4} \;\muas \;,
\label{thetas2}
\eeq
for a wind-like medium, where ${\cal E}_0 = 2\, E_0/(1 - \cos \theta_0)$ is the jet isotropic
equivalent energy and $A_* \equiv A/(3.0\times 10^{35}\; {\rm cm^{-1}})$ is $A$ normalized to 
the value corresponding to the ejection of $10^{-5}\,{\rm M_{\odot} yr^{-1}}$ at $10^3\,{\rm km\; 
s^{-1}}$). In equations (\ref{thetas0}) and (\ref{thetas2}) $D_A$ is the angular distance for 
redshift $z$, $t_d$ is the observer time measured in days, and the usual scaling $X_n = 10^{-n} X$ 
was used. 

 After $\Gamma$ drops below $\theta^{-1}$, which is roughly the onset time of the lateral spreading
dominated phase, the source size is $\theta_s = r \sin \theta$. This quantity and the arrival time 
of photons emitted from the jet edge are calculated numerically.

\subsection{Gravitational Microlensing}
\label{lens}

 The decay of the near infrared (NIR) and optical emission of 000301c exhibits a modest brightening
at 3--5 days, which appears to be achromatic in the optical (Berger \etal 2000, Bhargavi \& Cowsik 2000), 
although there is evidence for $R-K$ color evolution during it (Rhoads \& Fruchter 2001). Garnavich 
\etal (2000b) have proposed gravitational microlensing as an explanation for this feature, though 
other possibilities -- inhomogeneities in the external medium, delayed energy injection, departures 
of the electron distribution from a power-law -- cannot be ruled out given the lack of well 
sampled, simultaneous measurements in other frequency domains and of a detailed analysis of the
chromaticity of the brightening in all these scenarios.  
 
 We include microlensing in our modeling to assess its ability to explain the brightening of
the 000301c afterglow. Unlike the other scenarios mentioned above, lensing is ``external" to 
the afterglow modeling, as it does not modify the jet dynamics. Better yet, it introduces only 
two free parameters: \\
\hspace*{2mm} $1)\;\; \theta_{LS}$, the apparent separation between the lens and the center of 
                the GRB remnant and \\
\hspace*{2mm} $2)\;\; \theta_E$, the angular size of the Einstein disk, which depends on the
               lens mass and distance to the observer. 
Thus the total number of parameters of the afterglow model with lensing is 10. 

 As pointed out by Loeb \& Perna (1998), a time-varying magnification of afterglows results
when the expanding source (\eqs [\ref{thetas0}] and [\ref{thetas2}]) crosses the Einstein disk. 
The effect is enhanced by that the source has a non-uniform surface brightness distribution, 
with most of the optical emission coming from the outer part of the visible disk. These two 
factors are taken into account in our calculations by integrating over the jet surface the flux 
received from each infinitesimal patch multiplied by the magnification factor
\beq
 \mu = \frac{u^2+2}{u\sqrt{u^2+4}} \;,
\eeq
where $u$ is the angular separation (on the sky) between the patch and the lens, measured in 
units of $\theta_E$.

 We note some important differences between the model used by Garnavich \etal (2000b) and ours.
We calculate the afterglow emission from the jet dynamics instead of approximating it with a 
smoothed broken power-law. The source surface brightness distribution is implicitly taken into 
account by our integration of the afterglow emission. Thus we do not approximate the afterglow 
image as a ring, with zero brightness inside the ring. Garnavich \etal (2000b) used a free 
normalization factor for the afterglow emission in each of the 7 observational bands included 
in their fit. This allows a rather excessive freedom to their model, given that the afterglow 
spectrum in NIR--optical is a power-law, thus the normalizing factor at one frequency determines 
the factors for all other bands. All these differences point to that our model has significantly 
less freedom than that of Garnavich \etal (2000b).

\section{Spectral/Temporal Features of the 000301c Afterglow}

 The afterglow data which we model are taken from Bhargavi \& Cowsik (2000), Berger \etal (2000),
Masetti \etal (2000), Sagar \etal (2000), Jensen \etal (2001), and Rhoads \& Fruchter (2001). 
Some of these articles list measurements reported in GCN Circulars by Bernabei \etal (2000), 
Gal-Yam \etal (2000), Garnavich \etal (2000a), Halpern \etal (2000), Kobayashi \etal (2000), 
Mujica \etal (2000), and Veillet \& Boer (2000). We have excluded the earliest $R$-band magnitude 
reported by Sagar \etal (2000), which is inconsistent at the $6\sigma$ level with an almost 
simultaneous measurement by Bhargavi \& Cowsik (2000), and the $UV$ flux reported by Smette 
\etal (2001), which could be affected by intergalactic Ly$\alpha$ absorption. We have averaged 
measurements separated by less than 0.01 day, whether they were independent observations or the 
same data analyzed and reported by different workers. The final data set contains 140 points.

 The NIR and optical magnitudes have been converted to fluxes using the photometric zero points
published by Campins, Rieke, \& Lebofsky (1985), and Fukugita, Shimasaku \& Ichikawa (1995).
To account for magnitude-to-flux conversion uncertainty we add 5\% in quadrature to the 
reported uncertainties, noting that most of these uncertainties are equal to or exceed 5\%.
The NIR and optical fluxes have corrected for Galactic extinction with $E(B-V)=0.053$ 
(\eg Rhoads \& Fruchter 2001), but no correction was made for a possible reddening in the
afterglow's host galaxy. 

 Throughout this work we denote by $\alpha$ and $\beta$ the indices of the power-law afterglow
light-curve and spectrum:
\beq
 F_\nu (t) \propto \nu^{-\beta} t^{-\alpha} \;.
\label{plaw}
\eeq
Over various ranges in frequency and times $\alpha$ and $\beta$ are constants.

\subsection{Spectral Slopes}
\label{spectrum}

 For 000301c, the measurements made around $t=4.3$ day span the widest frequency range.
At the same time the optical emission exhibits a brightening. Figure 1 shows the afterglow 
spectrum at $t=5.0$ day, when the optical bump seems to have faded, obtained from the data 
closest to this time, using interpolations or extrapolations where necessary. 

 The {\sl radio} spectrum of 000301c has a slope $\beta_r = -0.75 \pm 0.15$, between the 
values expected for optically thin ($\beta_r = -1/3$) and thick ($\beta_r = -2)$ emission, 
thus the absorption frequency $\nu_a$ must be around 10 GHz at 5 days.

 Using the 250 GHz measurements reported by Berger \etal (2000) and the $K$-band data of 
Rhoads \& Fruchter (2001), we find that the {\sl millimeter--NIR} spectral slope $\beta_{mK}$ 
is
\beq
 \beta_{mK} = 0.67 \pm 0.04 \quad {\rm at} \quad t = 5 \; {\rm d} \;.
\label{betamK}
\eeq

 Rhoads \& Fruchter (2001) found that, after correction for Galactic extinction only, the 
spectral slope between the $K$- and $R$-bands is (see their Table 3) 
\beq
 \beta_{KR} = \left\{  \begin{array}{ll}
  0.72 \pm 0.04 & t=2.1 \; {\rm d} \\ 
  0.91 \pm 0.04 & t=3.0 \; {\rm d} \\ 
  0.99 \pm 0.15 & t=5.0 \; {\rm d} \\
  0.69 \pm 0.10 & t=7.6 \; {\rm d}  \end{array} \right. \;,
\label{betaKR}
\eeq
while at an earlier time, $t=1.8$ day, the $R-K$ color presented in their figure 2 gives 
$\beta_{KR} = 0.47 \pm 0.11$. Rhoads \& Fruchter (2001) find that, in order to reconcile
the data with no evolution of $\beta_{KR}$, a systematic error of 0.08 mag has to be 
added in quadrature to the $K$ and $R$-band fluxes, and conclude that at least some of
$K-R$ color evolution is real.

 From spectroscopic observations, Feng \etal (2000) found that the {\sl optical} 
spectral slope is $\beta_O \sim 1.1$ at $t=2$ day, while Jensen \etal (2001) arrived at 
\beq
 \beta_O =  \left\{  \begin{array}{ll}
 1.15 \pm 0.26 & t=4.0 \; {\rm d} \\
 1.43 \pm 0.28 & t=5.0 \; {\rm d}  \end{array} \right. \;.
\label{betaO}
\eeq
Equations (\ref{betaKR}) and (\ref{betaO}) lead to
\beq
 \beta_O - \beta_{KR} =  \left\{  \begin{array}{ll}
  0.23 \pm 0.26  & t\sim 4\; {\rm d} \\ 
  0.44 \pm 0.32  & t=5  \; {\rm d} \end{array} \right. \;.
\label{dbetaKO}
\eeq
The above results provide only weak evidence for a curved (even after dereddening for Galactic 
extinction), softening afterglow spectrum. Jensen \etal (2001) and Rhoads \& Fruchter (2001) 
attribute this departure from the power-law spectrum expected in the fireball model to extinction 
in the host galaxy. We note that softening of the NIR--optical spectrum, if real, suggests an 
intrinsic feature of the afterglow, as caused by a spectral break whose frequency decreases,
passing through the optical domain. Given that the fastest possible decrease of the cooling frequency 
is $\nu_c \propto t^{-1/2}$ (see Panaitescu \& Kumar 2001) and that the steepening $\Delta \beta 
= 1/2$ of the spectrum across $\nu_c$ is smooth, it seems rather unlikely that the passage of
$\nu_c$ would produce a detectable afterglow softening over a short timescale. This suggests 
that the evolving break is the $\nu_*$ associated with the high end of the electron distribution 
($\gamma_*$), which evolves as fast as $\nu_i$ and can be sharper than the $\nu_c$-break.  

 It is also possible that the chromatic evolution of the afterglow between 2 and 5 days is caused
by the same mechanism that yields the optical brightening seen at $\sim 3.5$ days. Most likely this 
excludes microlensing as the brightening mechanism, as lensing should not produce a chromatic 
afterglow behavior over less than a decade in frequency (note that, in general, microlensing of 
afterglows leaves chromatic signatures because the surface brightness profile of the source is 
wavelength-dependent).

\subsection{Light-Curve Decay Indices}
\label{index}

 The 8.46 GHz {\sl radio} emission of the 000301c afterglow, shown in Figure 2, exhibits a 
power-law decline with 
\beq
 \alpha_r = 1.0 \pm 0.2 \quad {\rm at} \quad t \simg 30\; {\rm d} \;.
\label{alphar}
\eeq
A steeper and highly uncertain decline, between $t^{-1}$ and $t^{-2.5}$, is seen in the 4.86 
GHz emission. Due to the curvature of the emitting surface, relativistic jets cannot produce 
significantly different decay indices at frequencies that are a factor of only two apart, so 
we presume that the light-curve indices at 4.86 GHz and 8.46 GHz are in fact the same and that 
the discrepancy can be resolved by interstellar scintillation.

 The {\sl millimeter} decay index at 250 GHz is not well determined by the data, nevertheless 
the $2\sigma$ upper limits shown in Figure 2 indicate that $\alpha_m > 0.7$. This shows 
that after few days the injection frequency $\nu_i$ is below $\sim 100$ GHz, unless the external 
medium is wind-like and the ordering of frequencies is $\nu_a < 250\; {\rm GHz} < \nu_c < \nu_i$, 
in which case it is analytically expected that $\alpha_m = 2/3$ (Panaitescu \& Kumar 2000), 
marginally consistent with the observations.

 The NIR and {\sl optical} emission of 000301c exhibited a break at few days. Using a smoothed
dual power-law fit, Rhoads \& Fruchter (2001) found a steepening from an asymptotic $\alpha_{K1} 
= 0.09$ at early times to $\alpha_{K2} = 2.26$ at late times, in the $K$ band, and from 
$\alpha_{R1} = 0.69$ to $\alpha_{R2} = 2.77$ in the $R$ band. Using a broken power-law 
approximation, Jensen \etal (2001) found $\alpha_{O1} = 0.73 \pm 0.27$ and $\alpha_{O2} = 2.67 
\pm 0.51$. At late times, when the effect of microlensing should be negligible, Garnavich \etal 
(2000b) found $\alpha_{O2} = 2.9$, close to the value reported by Sagar \etal (2000):
$\alpha_{O2} = 2.97 \pm 0.04$. With a more complex fitting function and without the last two 
$R$ magnitudes reported by Fruchter \etal (2000), which constrain the late time decay index, 
Bhargavi \& Cowsik (2000) found $\alpha_{R1} = 0.70 \pm 0.07$ and $\alpha_{R2} = 2.44 \pm 0.29$.
Thus the asymptotic indices of the optical decay of 000301c are
\beq
  \alpha_{R1} = 0.70 \pm 0.07 \;,\quad \alpha_{R2} \geq 2.7 \;, 
\label{alphaR}
\eeq
implying that the decline of optical emission of this afterglow steepened by 
\beq
  \Delta \alpha_R = \alpha_{R2} - \alpha_{R1} \geq 1.9 
\label{dalphaR}
\eeq
over about one decade in observer time. 

 For a homogeneous external medium, the afterglow decay index is expected to increase by 3/4 
when the jet edge becomes visible (\ie the jet Lorentz factor drops below the inverse of
its aperture, $\theta$), with an extra steepening of at most $p/4$ arising from the jet 
dynamics.  It is thus hard to see how collimation of ejecta could account alone for the 
magnitude $\Delta \alpha_R$ of the optical break of 000301c, unless the afterglow decay before 
3 days is significantly flattened by the brightening mechanism we see at 3--5 days, so that 
the observed decay index $\alpha_{R1}$ is smaller than that the afterglow would have if this 
mechanism were not present. Given that it takes more than two decades in time to a jet expanding 
into an $r^{-2}$ medium to yield about 90\% of the full break magnitude (Kumar \& Panaitescu 
2000), the $\Delta \alpha_R$ above rules out a model with a wind-like medium and an afterglow 
steepening due only to collimated ejecta.

\section{Models for the 000301c Afterglow}
\label{models}

  We start with a model where the high energy spectral break $\nu_*$ is well above the 
optical domain and the light-curve steepening is due solely to the collimation of ejecta.
In this case the external medium must be homogeneous, so that the light-curve steepening 
at the time $t_j$, when the jet edge becomes visible, is as fast as possible. 

 If the cooling frequency were below the NIR-optical domain, then $\beta_{KR} = p/2$, thus the 
observed slope at 5 days (\eq [\ref{betaKR}]), after the optical brightening, would require 
$p = 1.98 \pm 0.30$. Then the decay index of the optical emission at $t > t_j$ would approach 
asymptotically $\alpha_{R2} \sim p \siml 2.3$, and would be shallower than observed 
(\eq [\ref{alphaR}]). Therefore the cooling frequency should be above the NIR--optical domain. 

 At few days, the injection frequency $\nu_i$ must be below the millimeter range, to explain 
the decay of the 250 GHz emission (\S\ref{index}, Figure 2). Then $\beta_{mK} = (p-1)/2$ and 
the observations at 5 days (\eq [\ref{betamK}]) lead to $p = 2.34 \pm 0.08$, which is still 
below $\alpha_{R2}$. Therefore it may be difficult to reconcile within the jet model the 
spectrum and the late time optical decay of 000301c. 

 The best fit obtained with a jet model having the cooling frequency above optical and a 
homogeneous external medium is shown in Figure 2. Its $\chi^2 = 680$ for 132 degrees of freedom 
(df) is large in part because microlensing does not produce the brightening seen around 3.5 days,
for reasons discussed below, the 28 NIR/optical measurements between 3.0 and 4.3 days yielding 
$\Delta \chi^2 = 210$. To test the ability of only the jet model to account for the general 
features of 000301c, we eliminate the 3.0--4.3 days data and microlensing from the modeling. 
The new best fit has $\chi^2 = 480$ for 98 df (see Figure 2). 

 One reason for which the new $\chi^2$ is still large is that a relativistic jet yields the
same $t^{-p}$ decay after $t_j$ at all frequencies above $\nu_i$. However, in 000301, the
late time decay indices of the 8.46 GHz and $R$-band light-curves differ significantly. 
For the jet models shown in Figure 2, the 14 measurements made at 8.46 GHz after 30 days 
yield $\Delta \chi^2 = 68$. Berger \etal (2000) explain this discrepancy (see their Figure 2) 
between the jet model and the radio emission of 000301c as the result of interstellar 
scintillation. Even if ISS were strong enough to accommodate the observations, the radio 
data should have been randomly scattered around the model light-curve, both having the same 
trend, which is unlike the behaviors illustrated in Figure 2. 

 The results of Figure 2 point to yet another reason for the large $\chi^2$ we obtain: the 
optical steepening of 000301c is too fast and too large to be explained with a jet break, 
as was expected based on analytical considerations (\S\ref{index}). For this reason we turn now 
to a ``modified jet model", where the passage through the optical of the $\nu_*$ break associated 
with the steepening of the electron distribution at $\gamma_*$ yields a light-curve steepening. 
As noted in \S\ref{spectrum}, the passage of a spectral break is suggested by the softening 
of the NIR--optical spectrum (\eq [\ref{betamK}]). The time and magnitude of the light-curve break 
caused by the $\nu_*$-crossing depend on the location of $\gamma_*$ and on the steepening of the 
electron distribution at $\gamma_*$, \ie the parameters $\epsilon$ and $\delta p$ introduced in 
\S\ref{synchro}. 

 First note that, within the jet model, the quasi-constant 8.46 GHz emission prior to 30 days
has two possible explanations: \\ 
\hspace*{2mm}
 $1)$ the ejecta are collimated, the jet edge becoming visible at few days, so that 
  $F_\nu \propto t^{-1/3}$ before the passage of $\nu_i$ and $F_\nu \propto t^{-p}$ after that. 
  These analytical results (Rhoads 1999) are not very accurate for moderately relativistic jets
  ($\Gamma < 10$), where we find numerically smaller decay indices. Nevertheless a hard electron 
  distribution, with $p \siml 1.5$, is required by the observed $\alpha_r \sim 1$. 
  Since the observed $\beta_{mK}$ (\eq [\ref{betamK}]) is larger than $(p-1)/2$, the cooling 
  break must be below NIR. The external medium can be either homogeneous or wind-like. \\
\hspace*{2mm} 
 $2)$ the external medium is a wind and the ejecta spherical, the radio emission evolving 
  as $F_\nu \approx const$ before $\nu_i = \nu$ and as $F_\nu \propto t^{-(3p-1)/4}$ thereafter. 
  The observed $\alpha_r$ (\eq [\ref{alphar}]) requires that $p = 1.7 \pm 0.3$. As in the first 
  case, $\beta_{mK} > (p-1)/2$ implies that $\nu_c$ is below NIR. Given that the jet break is 
  very smooth if the external medium is wind-like, this case is not readily distinguishable
  from a jet interacting with a wind.

 Figure 3 shows the best fits we obtain with a modified jet interacting with a {\sl homogeneous} 
medium, with and without lensing and the 3.0--4.3 day data. In the latter case, the resulting fit 
has $\chi^2 = 120$ for 96 df, being marginally acceptable. Note that, in either case, the modified
jet yields a significantly better fit than the "traditional" jet model with $p > 2$ and optical
steepening caused by collimation. The best fits obtained with a modified jet and
a {\sl wind-like} medium are shown in Figure 4. Without microlensing and the 3.0--4.3 days data, 
it has $\chi^2 = 140$ for 96 df, being unacceptable. The light-curves shown in Figures
3 and 4 prove that, for either type of external medium, microlensing does not yield an afterglow
brightening but only a flattening, for the reasons discussed below.

 As found by many researchers (Waxman 1997, Panaitescu \& M\'esz\'aros 1998, Sari 1998, Granot, 
Piran, \& Sari 1999), at high frequencies, a spherical, relativistic fireball appears as a disk 
whose brightness surface increases toward the edge. An increasing magnification through lensing 
requires that the source is initially outside the Einstein disk. The peak magnification is obtained 
when, due to the source expansion (\eqs [\ref{thetas0}] and [\ref{thetas2}]), its edge crosses the 
Einstein disk. A significant magnification for a short duration (1--2 days for 000301c)
constrains the size of the Einstein disk in two ways: it cannot be much smaller than the source 
size at the time of maximum lensing, as in this case the magnification would be weak, but it cannot 
be larger than it either, because the lensing event would last longer than the time when the
maximum magnification is produced (3--4 days for 00301c). For the jet models with microlensing
presented in Figure 2--4, the evolution of the source size and of the surface brightness distribution
are such that the best fits are obtained if, at the time of maximum magnification ($\sim 3.5$ days), 
the afterglow apparent size is twice larger than the Einstein disk. For this source--lens geometry,
the source is too large for microlensing to yield a peaked light-curve. 

 Garnavich \etal (2000b) have obtained peaked lensed light-curves by approximating the source 
image with an annulus, whose fractional width (16\%) was obtained by fitting the data. Granot \& Loeb 
(2001) have calculated microlensed light-curves using the surface brightness distribution of a 
spherical, relativistic afterglow, and have found a peak magnification factor $\sim 2$, supporting 
the microlensing model of Garnavich \etal (2000) for 000301c. We note, however, that in 000301c,
the optical brightening occurs at times when jet effects are important, thus the source geometry
and dynamics are not those expected for a spherical fireball. According to Ioka \& Nakamura (2001), 
the surface brightness distribution of a spreading jet is rather uniform. We also note that, at the 
time of optical brightening, the jets of Figure 3 and 4 are not very relativistic ($\Gamma \sim$ 
several), which makes the surface brightness distribution even more uniform. Therefore, in our
modeling of 000301c, jet effects and the mildly relativistic motion of the source lead to an
afterglow whose image at optical frequencies is not like the narrow ring envisaged by Garnavich
\etal (2000b), but more like a disk.

\section{Conclusions}

 The analysis presented in \S\ref{models} shows that the ``traditional" jet model cannot 
reconcile the late time decay indices of the 8.5 GHz and $R$-band light-curves of 000301c, 
and cannot account for the magnitude of the steepening exhibited by decay of its optical 
emission. We attribute the latter to the passage of a spectral break $\nu_*$ associated with 
a steepening of the power-law electron distribution at high Lorentz factors. This also 
provides a natural explanation for the curvature of the optical spectrum observed by Jensen 
\etal (2001) without the help of host extinction, and for the possible softening of the 
NIR--optical spectrum implied by the results of Jensen \etal (2001) and Rhoads \& Fruchter 
(2001), at about the same time when the steepening of the optical fall-off of 000301c occurred. 
Note that a chromatic evolution of the afterglow emission over less than a decade in frequency 
cannot be due to microlensing or to jet effects.

 A break in the electron distribution was also used by Li \& Chevalier (2001) to model the
emission of 000301c in the framework of spherical ejecta interacting with a wind-like external
medium, and by Panaitescu \& Kumar (2001) for the afterglow of GRB 991216, where its was
found that a jet model with a broken power-law injected electron distribution accommodates 
the data better than the simple jet model.
Although the optical light-curve breaks shown in Figures 3 and 4 are mostly due to the 
$\nu_*$-passage, the collimation of ejecta is still present in the model, to obtain consistency
between the hard electron index $p \siml 1.5$ and the $t^{-1}$ late radio decay.

 We find marginally acceptable fits only for a jet interacting with a homogeneous medium, and
after excluding the 3.0--4.3 days data, as we could not model the brightening of 000301c during 
this time using microlensing. From the variation of $\chi^2$ around its minimum, we arrive at the 
following 95\% confidence level intervals for a single parameter:
\bdm
  E_0 = (1.8 - 3.6) \times 10^{50} \; {\rm erg} \;, \quad
  \theta_0 = 10.8 - 14.6  \; {\rm deg} \;,
\edm
\bdm
  n = 13 - 38 \; {\rm cm^{-3}} \;, \quad \epsmag = 0.032 - 0.15 \;,
\edm
\bdm
  \epsel = 0.030 - 0.065 \;, \quad p = 1.41 - 1.59 \;,
\edm
\bdm
  \epsilon = 0.34 - 0.54 \;, \quad \delta p = 0.9 - 1.3 \;.
\edm
These results are given {\sl only} to show the statistical uncertainties of the model parameters, 
and {\sl should not} be considered absolute limits. For such limits one would need to know the 
``model error bars", which may be larger than the above uncertainties. 
 
 Consistency between the expected decay of jet light-curves and that seen in 000301c at 8.5 GHz 
after 30 days requires a hard electron distribution, with $p \siml 1.6$, similar to the indices 
obtained by Malkov (1999) for Fermi acceleration in the limit when particles acquire a significant 
fraction of the shock energy, and by Ellison, Jones, \& Reynolds (1990) and Baring (2000) for 
particle diffusion driven by large angle scatterings. 

 We note that the time when the afterglow steepening is seen constrains the $\nu_*$ break frequency 
and determines the electron fractional energy $\epsilon$. In light of this, is interesting to note 
that for 000301c we find an electron fractional energy close to equipartition, providing thus a 
natural reason for steepening of the electron distribution at high energies. 

 The model lensed light-curves we obtain show that microlensing by can yield a significant 
magnification but does not provide a good fit to the observations, failing to produce a peak 
at maximum lensing. The reason is that the afterglow surface brightness is not sufficiently 
concentrated toward the edge of the visible disk to yield a light-curve peak when this edge 
crosses the Einstein disk of the lens. We emphasize that this conclusion is subject to the 
correctness of the assumptions made in our treatment of the jet model. 

 Evidently, we do not rule out microlensing of 000301c within the framework of other afterglow 
models (\eg non-uniform jets). An inhomogeneity in the external medium or a pile-up of electrons 
(Protheroe \& Stanev 1999) at the $\gamma_*$ break, above which electrons are inefficiently accelerated, 
could in principle explain the brightening that preceded the steepening of the optical emission
of 000301c.

\acknowledgments{It is a pleasure to thank Pawan Kumar (IAS) for useful suggestions and 
                 comments on this work}

\clearpage

\begin{figure*}
\centerline{\psfig{figure=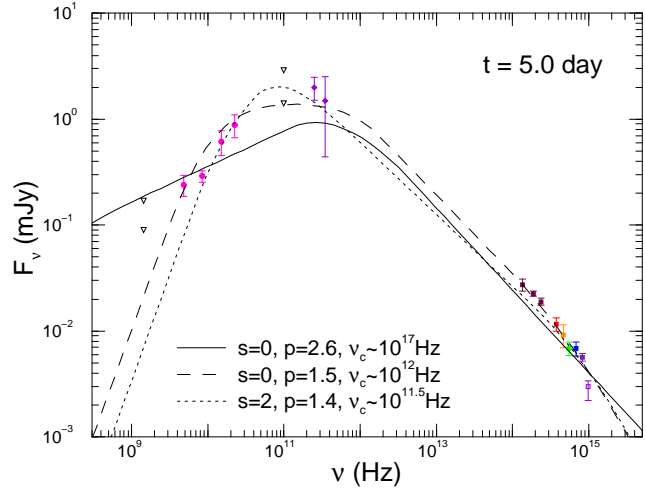}}
\figcaption{
 Spectrum of the 000301c afterglow at $t=5.0$ day constructed from the measurements closest 
  to this epoch and linear interpolations at low frequencies or using the fit of Rhoads \& 
  Fruchter (2001) in the NIR-optical.
 Optical fluxes are corrected for Galactic extinction of $E(B-V) = 0.053$.
 Downward triangles indicate $1\sigma$ and $2\sigma$ upper limits (note that Berger \etal
  2000 report a flux of $2.85 \pm 0.95$ mJy at 100 GHz an 4.26 days; the upper limits shown 
  here are their measurement at 5.09 days). The low flux at $10^{15}$ Hz, representing the 
  $UV$ measurement of Smette \etal (2001), could be due to intergalactic $Ly\alpha$ absorption. 
 The {\sl solid, dashed} and {\sl dotted} lines represents the spectra obtained with the jet 
  models described in Figures 2, 3, and 4, respectively.
}
\end{figure*}

\begin{figure*}
\centerline{\psfig{figure=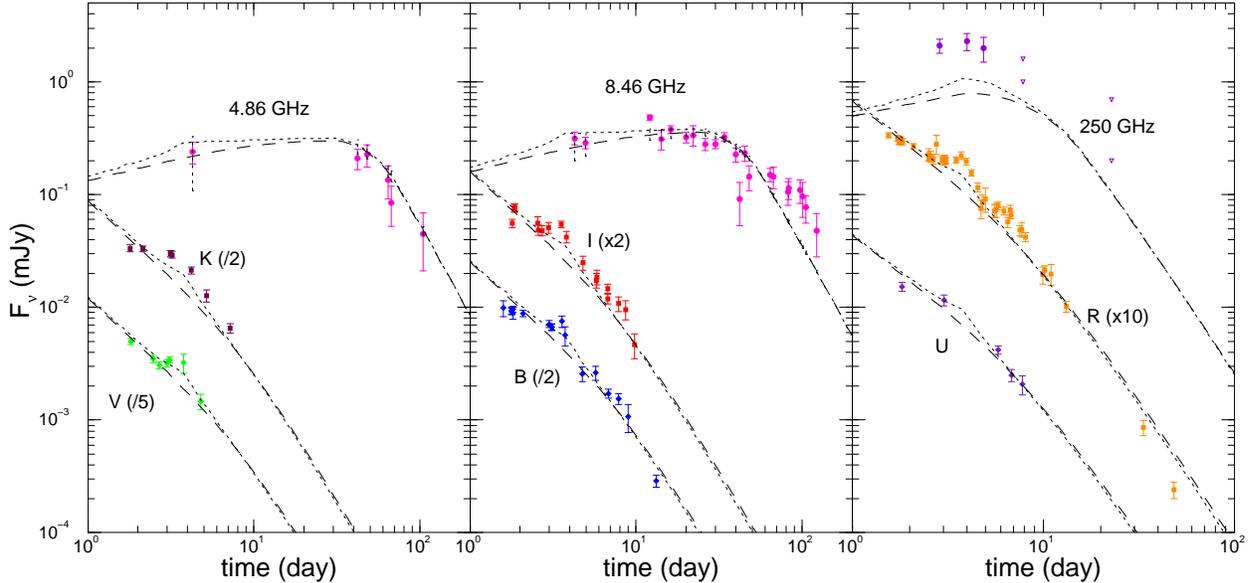}}
\figcaption{
  Best fit models for the 000301c afterglow with a {\sl soft} electron distribution, 
  {\sl homogeneous} external medium and cooling frequency above optical.
  Triangles show $1\sigma$ and $2\sigma$ upper limits.
  {\sl Dotted lines:} jet model with parameters (\S\ref{synchro}) 
   $E_0 = 1.9 \times 10^{51}\;{\rm erg}$, $\theta_0 = 3.1\deg$, $n = 0.015 \cm3$,
   $\epsel = 0.050$, $\epsmag = 4.1 \times 10^{-4}$, $p = 2.56$,
   lens parameters $\theta_{LS} = 4.5 \muas$ and $\theta_E = 2.7 \muas$,
   and $\chi^2 = 680$ for 132 df.
  {\sl Dashed lines:} jet model without microlensing, parameters
   $E_0 = 1.8 \times 10^{51}\;{\rm erg}$, $\theta_0 = 2.8\deg$, $n = 0.011 \cm3$,
   $\epsel = 0.047$, $\epsmag = 4.4 \times 10^{-4}$, $p = 2.53$, 
  and $\chi^2 = 480$ for 98 df, excluding the 3.0--4.3 day data.
  Vertical dotted segments illustrate the modulation amplitude due to interstellar scattering.
  Optical light-curves have been displaced vertically by the indicated factors.
  The afterglow is at redshift $z=2.03$ (Castro \etal 2000, Smette \etal 2001).
  We assumed $H_0 = 65\; {\rm km \, s^{-1} Mpc^{-1}}$, $q_0 = 0.1$, and $\Lambda = 0$.
}
\end{figure*}

\begin{figure*}
\centerline{\psfig{figure=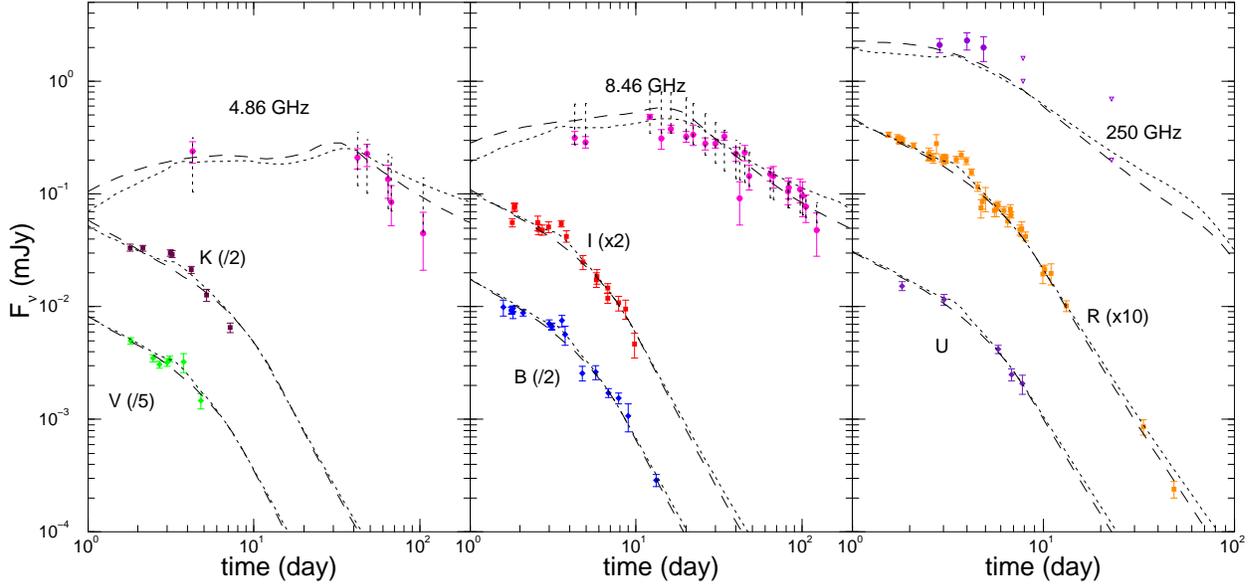}}
\figcaption{
  Best fit models with a {\sl hard} electron distribution, 
   cooling frequency below optical, and {\sl homogeneous} external medium.
  The jet model shown with {\sl dotted lines} includes microlensing, all data, and
   has $\chi^2 = 225$ for 130 df,
   $E_0 = 2.0 \times 10^{50}\;{\rm erg}$, $\theta_0 = 13.6\deg$, $n = 15 \cm3$, 
   $\epsmag = 0.18$, $\epsel = 0.045$, $p = 1.33$, $\epsilon = 0.33$, $\delta p = 1.1$,
   $\theta_{LS} = 0.90 \muas$ and $\theta_E = 0.38 \muas$.
  {\sl Dashed lines} are for a jet model without microlensing and the 3.0--4.3 day 
   measurements, and has $\chi^2 = 120$ for 96 df,
   $E_0 = 2.7 \times 10^{50}\;{\rm erg}$, $\theta_0 = 11.1\deg$, $n = 17 \cm3$, 
   $\epsmag = 0.072$, $\epsel = 0.037$, $p = 1.49$, $\epsilon = 0.31$, $\delta p = 1.1$.
  The vertical lines show the interstellar scintillation amplitude.
}
\end{figure*}

\begin{figure*}
\centerline{\psfig{figure=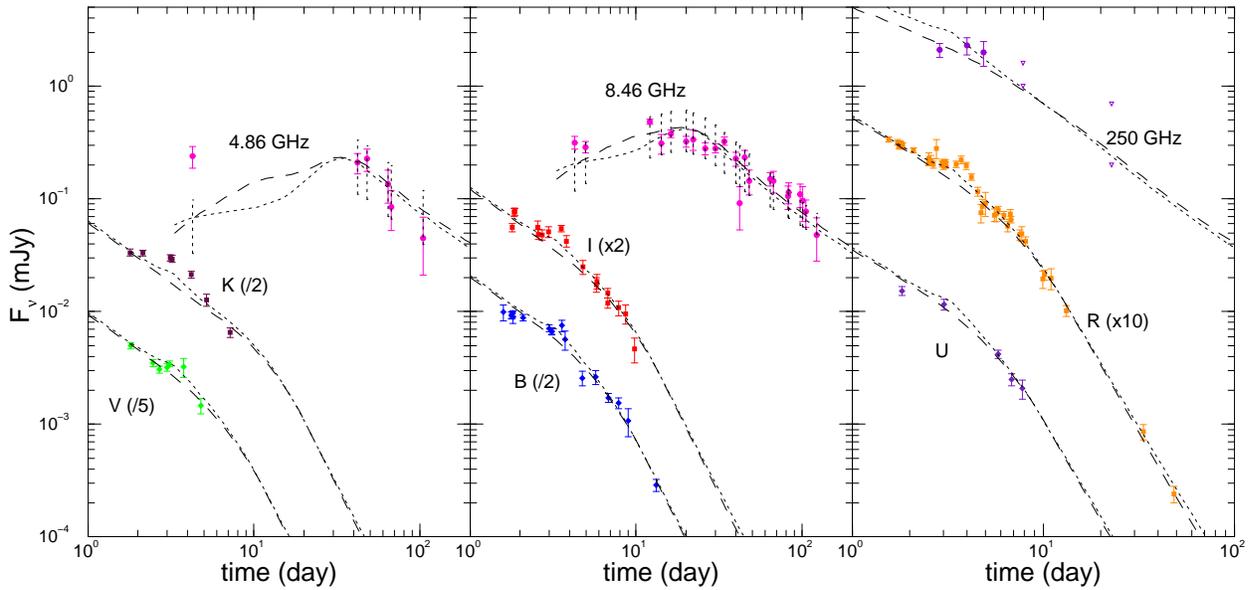}}
\figcaption{
  Best fit models with a {\sl hard} electron distribution,
   cooling frequency below optical, and {\sl wind-like} external medium.
  {\sl Dotted lines} represent the model with microlensing, all data, 
  $\chi^2 = 330$ for 130 df, and parameters
   $E_0 = 1.5 \times 10^{50}\;{\rm erg}$, $\theta_0 = 7.8\deg$, $A_* = 0.94$,
   $\epsmag = 0.25$, $\epsel = 0.022$, $p = 1.36$, $\epsilon = 0.32$, $\delta p = 1.6$,
    $\theta_{LS} = 0.62 \muas$ and $\theta_E = 0.29 \muas$.
  {\sl Dashed lines} are for a model without microlensing, excluding the 3.0--4.3 day data,
   and has $\chi^2 = 140$ for 96 df,
   $E_0 = 1.9 \times 10^{50}\;{\rm erg}$, $\theta_0 = 10.6\deg$, $A_* = 0.90$, 
   $\epsmag = 0.22$, $\epsel = 0.047$, $p = 1.36$, $\epsilon = 0.47$, $\delta p = 1.7$.
}
\end{figure*}

\end{document}